\documentclass[aps, pra, twocolumn, showpacs]{revtex4}
\usepackage{epsfig}
\usepackage{graphicx}
\usepackage{amsmath}
\usepackage{amsthm}
\usepackage{amssymb}
\usepackage{amsbsy}
\usepackage{amsfonts}
\usepackage{dsfont}
\usepackage{color}

\newcommand{\beq}{\begin{eqnarray}}
\newcommand{\eeq}{\end{eqnarray}}
\newcommand\bsy{\mathbf}
\newcommand\trace{\mathop\mathrm{Tr}\nolimits}

\newcommand\re{\mathop\mathrm{Re}}
\newcommand\im{\mathop\mathrm{Im}}

\def\rank{\mathop\mathrm{rank}}
\def\kb{k_\mathrm{B}}

\begin{document}
\title{Numerical evaluation of convex-roof entanglement measures\\ with applications to spin rings}
\author{Beat R\"othlisberger, J\"org Lehmann, and Daniel Loss}
\affiliation{Department of Physics, University of
Basel, Klingelbergstrasse 82, CH-4056 Basel, Switzerland}
\date{May 19, 2009}
\begin{abstract}
We present two ready-to-use numerical algorithms to evaluate
convex-roof extensions of arbitrary pure-state entanglement
monotones. Their implementation leaves the user merely with the task
of calculating derivatives of the respective pure-state measure. We
provide numerical tests of the algorithms and demonstrate their good
convergence properties. We further employ them in order to
investigate the entanglement in particular few-spins systems at
finite temperature. Namely, we consider ferromagnetic Heisenberg
exchange-coupled spin-$\frac{1}{2}$ rings subject to an
inhomogeneous in-plane field geometry obeying full rotational
symmetry around the axis perpendicular to the ring through its
center. We demonstrate that highly entangled states can be obtained
in these systems at sufficiently low temperatures and by tuning the
strength of a magnetic field configuration to an optimal value which
is identified numerically.
\end{abstract}

\maketitle

\section{Introduction}\label{sec:introduction}
Entanglement, one of the most intriguing features of quantum
mechanics \cite{Schrodinger1935, Einstein1935a}, is undoubtedly an
indispensable ingredient as a resource to any quantum computation or
quantum communication scheme~\cite{Nielsen2000a}. The ability to
(sometimes drastically) outperform classical computations using
multipartite quantum correlations has been demonstrated in various
theoretical proposals which by now have become well known standard
examples \cite{Deutsch1992, Cleve1998, Grover1996, Shor1997}. Due to
the rapid progress in the fields of quantum computation,
communication, and cryptography, both on the theoretical and the
experimental side, it has become a necessity to quantify and study
the production, manipulation and evolution of entangled states
theoretically.

However, this has turned out to be a rather difficult task, as the
dimension of the state space of a quantum system grows exponentially
with the number of qudits and thus permits the existence of highly
nontrivial quantum correlations between parties. While bipartite
entanglement is rather well understood (see, e.g.,
\cite{Plenio2007}), the study of multipartite states (with three or
more qudits) is an active field of research.

Several different approaches towards the study of entanglement
exist. Bell's original idea \cite{Bell1964} that certain quantum
states can exceed classically strict upper bounds on expressions of
correlators between measurement outcomes of different parties
sharing the same state has been widely extended and improved to
detect entanglement in a great variety of states. Entanglement
between photons persisting over large distances has been
demonstrated with the use of Bell-type inequalities (see, e.g., Ref.
\cite{Ursin2007} and references therein). Another more recent
approach is the concept of entanglement witnesses
\cite{Horodecki1996, Terhal2000a}. These are observables whose
expectation value is non-negative for separable states and negative
for some entangled states. Thirdly, the concept of entanglement
measures is focussing more on the quantification of entanglement: if
state $A$ has lower entanglement than state $B$, then $A$ cannot be
converted into $B$ by means of local operations and classical
communication. Remarkably, there exist interesting relations between
entanglement measures and Bell inequalities \cite{Emary2004} on the
one hand, and entanglement witnesses \cite{Guhne2007, Eisert2007} on
the other hand. In this work, we focus on the direct evaluation of
entanglement measures.

Among the many features one can demand of such a measure,
monotonicity is arguably the most important one: an entanglement
measure should be non-increasing under local operations and
classical communication (reflecting the fact that it is impossible
to create entanglement in a separable state by these means). A
measure exhibiting this property is called an entanglement monotone,
with prominent examples being, e.g., the entanglement of formation
\cite{Bennett1996}, the tangle \cite{Coffman2000}, the concurrence
\cite{Mintert2005a} or the measure by Meyer and Wallach
\cite{Meyer2002}. While one measure captures certain features of
some states especially well, other measures focus on different
aspects of different states.

Often, entanglement monotones are defined only for pure states and
are given as analytical expressions of the state's components in a
standard basis. Unfortunately, quantifying mixed-state entanglement
is more involved. This is somewhat intuitive, since the measure
needs to be capable of distinguishing quantum from classical
correlations. A manifestation of this difficulty is the fact that
the problem of determining whether a given density matrix is
separable or not is apparently very hard and has no known general
solution for an arbitrary number of subsystems with arbitrary
dimensions. The ability to study mixed-state entanglement is,
however, highly desirable since mixed-states appear naturally due to
various coupling mechanisms of the system under examination to its
environment. There exists a standard way to construct a mixed-state
entanglement monotone from a pure-state monotone, the so-called
convex-roof construction \cite{Uhlmann2000}, but the evaluation of
functions obtained in this way requires the solution of a rather
involved constrained optimization problem (see
Sec.~\ref{sec:convex-roof em}).

In this paper, we present two algorithms targeted at solving this
optimization problem numerically for any given convex-roof
entanglement measure. In principle, these algorithms can also be
applied to any optimization problem subjected to the same kind of
constraints. The first algorithm is an extension of a procedure
originally used to calculate the entanglement of formation
\cite{Audenaert2001}. It is a conjugate gradient method exploiting
the geometric structure of the nonlinear search space emerging from
the optimization constraint. The second algorithm is based on a real
parametrization of the search space, which allows one to carry out
the optimization problem in the more familiar Euclidean space using
standard techniques.

In the second part of the paper, we use these algorithms in order to
study the entanglement properties of a certain type of spin rings.
These systems form a generalization to $N$ qubits of our previous
study, where we had only considered the case $N = 3$
\cite{Rothlisberger2008}. In the presence of an isotropic and
ferromagnetic Heisenberg interaction and local in-plane magnetic
fields obeying a radial symmetry, it can be argued (see
Sec.~\ref{sec:spin system} and Ref.~\cite{Rothlisberger2008}) that
the ground state becomes a local unitary equivalent of an almost
perfect N-partite Greenberger-Horne-Zeilinger (GHZ) state
\cite{Greenberger1989}
\begin{equation}\label{GHZ state generalized}
|\mathrm{GHZ}_N^\pm\rangle =
\left(|{\uparrow\uparrow\ldots\uparrow}\rangle \pm
|{\downarrow\downarrow\ldots\downarrow}\rangle\right)/\sqrt{2}.
\end{equation}
Such a system could hence be used for the production of highly
entangled multipartite states merely by cooling it down to low
temperatures. One finds, however, that the energy splitting between
the ground and first excited state vanishes in the same limit as the
N-partite approximate GHZ states become perfect, namely for the
magnetic field strength going to zero. Therefore, in order to
quantitatively identify the magnetic field strengths yielding
maximal entanglement at finite temperature, one has to study the
system in terms of a suitable mixed-state entanglement measure.

The outline of the paper is as follows: In Sec.~\ref{sec:convex-roof
em} we review how the evaluation of a convex-roof entanglement
measure is related to a constrained optimization problem. We then
develop and describe the numerical algorithms capable of tackling
this problem in Sec.~\ref{sec:algorithms}. We also present some
benchmark tests, comparing our methods to another known algorithm.
In Sec.~\ref{sec:spin system}, we describe the spin rings mentioned
earlier and study their entanglement properties in terms of a
convex-roof entanglement measure evaluated using our algorithms. We
conclude our work in Sec.~\ref{sec:conclusion}.

\section{Convex-roof entanglement measures as constrained optimization
problems}\label{sec:convex-roof em}

Given a pure-state entanglement monotone $m$, the most reasonable
properties one can demand of a generalization of $m$ to mixed states
are that this generalization is itself an entanglement monotone, and
that it properly reduces to $m$ for pure states. A standard
procedure which achieves this is the so-called convex-roof
construction \cite{Uhlmann2000, Mintert2005}. Given a mixed state
$\rho$ acting on a Hilbert space $\mathcal{H}$ of finite dimension
$d$, it is defined as
\begin{equation}\label{convex-roof em}
M(\rho) = \inf_{\{p_i,
|\psi_i\rangle\}\in\mathfrak{D}(\rho)}\sum_{i}p_i m(|\psi_i\rangle),
\end{equation}
where
\begin{multline}
\mathfrak{D}(\rho) = \Bigl\{\left\{p_i,|\psi_i\rangle\right\}_{i =
1}^s, s \geq \rank\rho \;\big|\; \{|\psi_i\rangle\}_{i = 1}^s
\subset \mathcal{H}, \\ p_i \geq 0,\; \sum_{i = 1}^s p_i = 1, \;
\rho = \sum_{i=1}^s p_i |\psi_i\rangle\langle\psi_i| \Bigr\}
\end{multline}
is the set of all pure-state decompositions of $\rho$. Note that the
pure states $|\psi_i\rangle$ are understood to be normalized. The
numerical value of $M(\rho)$ is hence defined as an optimization
problem over the set $\mathfrak{D}(\rho)$.

In order to apply numerical algorithms to this problem,
$\mathfrak{D}(\rho)$ must be accessible in a parametric way. This
parametrization is well-known and is often referred to as the
Schr\"odinger-HJW theorem \cite{Hughston1993, Kirkpatrick2005},
which we briefly outline here for the sake of completeness.

Let $St(k, r)$ denote the set of all $k\times r$ matrices $U\in
\mathds{C}^{k\times r}$ with the property $U^\dag U = 1_{r\times
r}$, i.e., matrices with orthonormal column vectors (hence we have
$k\geq r$). The first part of the Schr\"odinger-HJW theorem states
that every $U\in St(k, r)$ yields a pure-state decomposition $\{p_i,
|\psi_i\rangle\}_{i = 1}^k \in \mathfrak{D}(\rho)$ of the density
matrix $\rho$ by the following construction. Let $\lambda_i$,
$|\chi_i\rangle$, $i = 1, \ldots, r = \rank\rho$ denote the
eigenvalues and corresponding normalized eigenvectors of $\rho$,
i.e.,
\begin{equation}
\rho = \sum_{i = 1}^r \lambda_i |\chi_i\rangle\langle\chi_i |.
\end{equation}
Note that we have $\lambda_i > 0$ since $\rho$ is a density matrix
and as such a positive semi-definite operator. Given a matrix $U\in
St(k, r)$, define the auxiliary states
\begin{equation}
|\tilde\psi_i\rangle = \sum_{j=1}^r
U_{ij}\sqrt{\lambda_j}|\chi_j\rangle, \qquad i = 1, \ldots, k.
\end{equation}
It is then readily checked that
\begin{gather}
p_i = \langle\tilde\psi_i|\tilde\psi_i\rangle, \\
|\psi_i\rangle = (1/\sqrt{p_i})|\tilde\psi_i\rangle
\end{gather}
is indeed a valid decomposition of $\rho$ into a convex sum of $k$
projectors.

The second part of the theorem states that for any given pure-state
decomposition $\{p_i, |\psi_i\rangle\}_{i = 1}^k$ of $\rho$, there
exists a $U\in St(k, r)$ realizing the decomposition by the above
construction. This guarantees that by searching over the set $St(k,
r)$ and obtaining the decompositions according to the
Schr\"odinger-HJW theorem, we do not `miss out' on any part of the
subset of $\mathfrak{D}(\rho)$ with a fixed number of states $k$.
The parameterization is thus complete, i.e., searching the infimum
over $St(k, r)$ is equivalent to searching over all decompositions
with fixed so-called cardinality $k$. This allows us to reformulate
the optimization problem Eq. \eqref{convex-roof em} as
\begin{equation}\label{general problem}
M(\rho) = \min_{k \geq r}\inf_{U\in St(k, r)} h(U),
\end{equation}
where $h(U)$ is the sum on the right-hand side of Eq.
\eqref{convex-roof em} obtained via the matrix $U$ from $\rho$,
i.e.,
\begin{equation}\label{convex sum}
h(U) = \sum_{i = 1}^k p_i(U) m(|\psi_i(U)\rangle).
\end{equation}
Note that we have dropped the $\rho$-dependence in the above
expressions, since $\rho$ is fixed within a particular calculation
and only the dependence of $h$ on $U$ is of relevance in the
following.

It is clear that in a numerical calculation only a finite number of
different values for $k$ can be investigated. However, it is also
intuitive to expect that for some large enough value of $k$,
increasing the latter even further has only marginal effects. In
fact, we have observed numerically that already $k = \rank\rho + 4$
yields very accurate results in all tests we have performed (also in
the ones presented in Sec.~\ref{sec:test cases}), and we have used
this choice throughout all numerical calculations within this work.
Note that for a fixed value of $k$, also all other decompositions
with cardinality smaller than $k$ are considered as well, since the
probabilities $p_i$ in the elements of $\mathfrak{D}(\rho)$ are
allowed to go to zero (with the convention that the corresponding
states $|\psi_i\rangle$ are then discarded).

Since the algorithms presented in the next section will both be
gradient-based, the derivatives of Eq. \eqref{convex sum} with
respect to the real and imaginary parts of $U$ evaluated at $U$ will
be required at some point. We state them here for the convenience of
the reader. They are given by
\begin{widetext}
\begin{align}
\frac{\partial h}{\partial \re U_{kl}} &=\label{partial h 1}
2\lambda_l\re(U_{kl})m(|\psi_k(U)\rangle) + \sum_{i = 1}^d\left[ \re
\phi^{(i)}_{R,kl} \left.\frac{\partial m}{\partial \re
\psi^{(i)}}\right|_{|\psi_k(U)\rangle} + \im \phi^{(i)}_{R,kl}
\left.\frac{\partial m}{\partial \im
\psi^{(i)}}\right|_{|\psi_k(U)\rangle}\right],\\
\frac{\partial h}{\partial \im U_{kl}} &=\label{partial h 2}
2\lambda_l\im(U_{kl})m(|\psi_k(U)\rangle) + \sum_{i = 1}^d\left[\re
\phi^{(i)}_{I,kl} \left.\frac{\partial m}{\partial \re
\psi^{(i)}}\right|_{|\psi_k(U)\rangle} + \im \phi^{(i)}_{I,kl}
\left.\frac{\partial m}{\partial \im
\psi^{(i)}}\right|_{|\psi_k(U)\rangle}\right],
\end{align}
\end{widetext}
where
\begin{align}
|\phi_{R, kl}(U)\rangle &=
\left[\sqrt{p_k(U)\lambda_l}|\chi_l\rangle -
\lambda_l\re(U_{kl})|\psi_k(U)\rangle\right],\\
|\phi_{I, kl}(U)\rangle &=
\left[\mathfrak{i}\sqrt{p_k(U)\lambda_l}|\chi_l\rangle -
\lambda_l\im(U_{kl})|\psi_k(U)\rangle\right],
\end{align}
and superscripts such as in $\psi^{(i)}$ denote the $i$th component
of the state $|\psi\rangle$ in an arbitrary but fixed basis.

As a last remark, we would like to point out that the constraint set
$St(k, r)$ is, in fact, a closed embedded submanifold of
$\mathds{C}^{k\times r}$, called the complex Stiefel manifold
\cite{Absil2008}. The geometric structure emerging thereof is
exploited in one of the two algorithms following shortly. The
dimension of the Stiefel mainfold is $\dim St(k, r) = 2kr - r^2$
\cite{Absil2008}. Since we have $k \geq r$, we can set $k = r + n$,
$n = 0, 1, \ldots$. The number of free parameters $N$ in the
optimization is thus $N = r^2 + 2nr$. Hence, $N$ grows linearly with
$n$, but quadratically with $r$. Numerical evaluation in larger
systems will thus be restricted to low-rank density matrices. The
flexibility of choosing $n$ is however less restricted. As mentioned
above, $n = 4$ already yields satisfying results.

\section{Numerical Algorithms}\label{sec:algorithms}

The study of optimization problems on matrix manifolds is a rather
new and still active field of research (see \cite{Edelman1998,
Absil2008} and references therein). Only recently, two ready-to-use
algorithms for minimization over the complex Stiefel manifold have
been presented \cite{Manton2002}. To our knowledge, these are the
only general purpose algorithms applicable to generic target
functions over $St(k, r)$ found in the literature. One is a steepest
descent-type method, the other one is of Newton-type. We will
compare the performance of the modified steepest descent algorithm,
as it is referred to in the original work, with the methods
presented in this section. We have found that our algorithms
generally show better convergence properties in the cases we have
examined.

We will, however, not make use of the modified Newton algorithm for
the following reasons. The second derivatives (as required by any
Newton-type algorithm) of the function $h(U)$ [Eq. \eqref{convex
sum}] are in general quite involved and their number grows
quadratically with the size of $U$. Hence, they are very expensive
to evaluate, even if one resorts to numerical finite differences.
Moreover, the good convergence properties of Newton-type methods may
only be expected in the very proximity of a local minimum. One
therefore first typically employs gradient-based techniques to
approach a minimum sufficiently enough. However, what `sufficiently
enough' means in a particular case is often not known beforehand. We
will later make use of a quasi-Newton algorithm, which approaches
local minima satisfyingly and shows strong convergence similar to
Newton methods automatically when being close enough to a minimum.

\subsection{Generalized Conjugate-Gradient Method}\label{sec:cg
method}

In Ref.~\cite{Audenaert2001} a conjugate-gradient algorithm on the
unitary group $U(k) = St(k, k)$ was presented. The goal there was to
calculate the entanglement of formation also for systems with
dimensions different from $2\times 2$ \cite{Wootters1998}. Here, we
extend this result by noting that the method is applicable to any
optimization problem on $St(k, k)$, particularly to the evaluation
of entanglement measures other than the entanglement of formation,
and we calculate the required general expression of the gradient of
$h(U)$.

Optimizing over $St(k, k)$ instead of $St(k, r)$ comes at the cost
of over-parameterizing the search space. When using this algorithm
to calculate convex-roof entanglement measures, we simply took into
account only the first $r$ columns of the matrix obtained at every
iteration. This is certainly an aspect one could improve upon in
future research.

The algorithm presented here is a conjugate gradient-type method,
meaning that instead of simply going downhill, i.e., in the
direction of steepest descent, previous search directions are taken
into account at the current iteration step. Once the search
direction $X_i$ at iteration step~$i$, a skew-Hermitian $k\times k$
matrix, is known, a line search along the geodesic $U_i
\exp({tX_i})$ is performed, where $U_i$ is the current iteration
point. In particular, one iteration step of the algorithm may be
described as follows \cite{Audenaert2001}:
\begin{enumerate}
\item Perform a line minimization, i.e., set
\begin{equation}
{t_{i+1} \leftarrow \arg\min_{t} h(U_i \exp{(t X_i)})}
\end{equation}
and set
\begin{equation}
U_{i + 1} \leftarrow U_i\exp{(t_{i+1} X_i)}.
\end{equation}

\item Compute the new gradient $G_{i+1}$ at $U_{i+1}$ and set
\begin{equation}
T \leftarrow \exp(t_{i+1}X_i/2)G_i\exp(-t_{i+1}X_i/2).
\end{equation}
$T$ is the gradient $G_i$ parallel-transported to the new point $U_{i+1}$.

\item Calculate the modified Polak-Ribi\`ere parameter
\begin{equation}
\gamma \leftarrow \frac{\langle G_{i+1} - T, G_{i+1}\rangle}{\langle G_i, G_i\rangle},
\end{equation}
where $\langle X, Y\rangle = \trace XY^\dag$.

\item Set the new search direction to
\begin{equation}
X_{i+1} \leftarrow -G_{i+1} + \gamma X_i.
\end{equation}

\item $i \leftarrow i + 1$.
\item Repeat from step 1 until convergence.

\end{enumerate}
The starting point $U_0$ can be chosen arbitrarily, and the initial
search direction is set to $X_0 = -G_0$. In order to find a good
approximation to the global minimum, one should restart the
procedure several times using random initial conditions. For the
line search in step 1, we utilized the derivative-free algorithm
\texttt{linmin} described in Ref.~\cite{Press1992}.

In the following, we calculate the general expression for the
gradient $G$ of the function $h$, evaluated at the point $U$ (we
drop iteration indices for simplicity). The gradient $G$ is defined
in terms of the directional derivative of $h$, namely as
\begin{equation}\label{grad start}
{\frac{dh(U^{(\varepsilon)} (X))}{d\varepsilon}}\bigg|_{\varepsilon
= 0} = \langle G, X\rangle,
\end{equation}
where $U^{(\varepsilon)}(X) = V\exp(\varepsilon X)$ is a geodesic on
$St(k, k)$ in direction $X$ (skew-Hermitian matrix) and passing
through $V$. The inner product is defined as in step 3 of the
algorithm. We will eventually read off the gradient $G$ from its
definition in Eq. \eqref{grad start}.

Treating $h(U)$ as a function of the real and imaginary matrix
elements of $U$, $\re U_{ik}$ and $\im U_{ik}$, respectively, we
have
\begin{equation}\label{grad chain}
\begin{split}
{\frac{dh(U^{(\varepsilon)} (X))}{d\varepsilon}}\bigg|_{\varepsilon = 0} &= \sum_{ik}\left(\frac{\partial h}{\partial\re U_{ik}}\bigg|_V \frac{\partial \re U^{(\varepsilon)}_{ik}}{\partial\varepsilon}\bigg|_{\epsilon = 0}\right. \\
&\quad \left. + \frac{\partial h}{\partial\im U_{ik}}\bigg|_V
\frac{\partial \im
U^{(\varepsilon)}_{ik}}{\partial\varepsilon}\bigg|_{\epsilon =
0}\right).
\end{split}
\end{equation}
The partial derivatives of $h$ with respect to $\re U_{ik}$ and $\im
U_{ik}$ have already been stated in Eqs. (\ref{partial h 1},
\ref{partial h 2}). Inserting the derivatives of
$U_{ik}^{(\varepsilon)}$ into Eq. \eqref{grad chain} and sorting all
terms with respect to $\re X$ and $\im X$, we obtain
\begin{equation}
{\frac{dh(U^{(\varepsilon)} (X))}{d\varepsilon}}\bigg|_{\varepsilon = 0}
= \sum_{kl}(A_{kl}\re X_{kl} + S_{kl}\im X_{kl}),
\end{equation}
where
\begin{align}
A_{kl} &= \sum_i \left(\frac{\partial h}{\partial\re U_{il}}\bigg|_V \re V_{ik} + \frac{\partial h}{\partial\im U_{il}}\bigg|_V \im V_{ik} \right), \\
S_{kl} &= \sum_i \left(\frac{\partial h}{\partial\im U_{il}}\bigg|_V
\re V_{ik} - \frac{\partial h}{\partial\re U_{il}}\bigg|_V \im
V_{ik} \right).
\end{align}
Taking into account the symmetry conditions on $X$ by using the
relations $\re X = (X - X^T)/2$ and $\im X = -\mathfrak{i}(X +
X^T)/2$ we further obtain
\begin{equation}
{\frac{dh(U_\varepsilon (X))}{d\varepsilon}}\bigg|_{\varepsilon = 0}
= \frac{1}{2}\sum_{kl}\big( (A_{kl} - A_{lk}) -\mathfrak{i}(S_{kl} +
S_{lk})\big)X_{kl}.
\end{equation}
By comparing this to the right-hand side of Eq.~\eqref{grad start},
i.e.,
\begin{equation}
\langle G, X\rangle = \trace GX^\dag = -\sum_{kl} G_{lk}X_{kl},
\end{equation}
we finally obtain the desired expression for the matrix elements of the gradient $G$,
\begin{equation}
G_{kl} = (A_{kl} - A_{lk})/2 + \mathfrak{i}(S_{kl} + S_{lk})/2.
\end{equation}
One readily sees that $G$ is skew-Hermitian, as required.

By this, we have completed the description of the conjugate gradient
algorithm capable of evaluating any convex-roof entanglement measure
presented in the form of Eq. \eqref{general problem}.

\subsection{Parametrization with Euler-Hurwitz angles}\label{sec:parameterization
method}

Here we present an alternative approach to optimization problems
over the Stiefel manifold $St(k, r)$. We will obtain a
parametrization of $St(k, r)$ in terms of a set of real numbers
which we will call Euler-Hurwitz angles, therefore unconstraining
the optimization problem and mapping it to Euclidean space, where
optimization problems have been investigated for much longer. We
will therefore be able to employ a standard algorithm to tackle the
transformed problem Eq. \eqref{general problem} \cite{Nocedal1999}.

The idea of parameterizing $St(k, r)$ is somewhat motivated by a
theorem known in classical mechanics, where it is stated that any
rotation in three-dimensional Euclidean space can be written as a
sequence of three elementary rotations described by three angles,
the Euler angles. In other words, any orthogonal $3\times 3$ matrix
is parameterized by three real numbers. It was already Euler himself
who generalized this idea to arbitrary $k \times k$ orthogonal
matrices \cite{Euler1987}, and Hurwitz \cite{Hurwitz1933} extended
the parametrization to unitary matrices. We remark that ideas in a
similar fashion to the ones promoted here have been used to
calculate an entanglement measure for Werner states
\cite{Terhal2002} but were not discussed in greater detail.

We now derive the parametrization of $St(k, r)$. Let $A\in St(k,
r)$. The basic idea is to generate zeroes in $A$ and bring it to
upper triangular form by applying so-called (complex) Givens
rotations $G_s(\vartheta, \varphi)$ \cite{Golub1996} to $A$ from the
left. The $k\times k$ matrices $G_s(\vartheta, \varphi)$, $s =
1,\ldots, k-1,$ are defined as
\begin{equation}
G_s^{i, j}(\vartheta, \varphi) = \begin{cases} e^{\mathfrak i\varphi}\cos\vartheta, & \mbox{if} \;\; i = j = s, \\
e^{-\mathfrak i\varphi}\sin\vartheta, & \mbox{if} \;\; i = s, j = s + 1,\\
-e^{\mathfrak i\varphi}\sin\vartheta, & \mbox{if} \;\; i = s + 1, j = s,\\
e^{-\mathfrak i\varphi}\cos\vartheta, & \mbox{if} \;\; i = s + 1, j = s + 1,\\
\delta_{ij},                 & \mbox{otherwise}.
\end{cases}
\end{equation}
Multiplying $A$ from the left with $G_s(\vartheta, \varphi)$, i.e.,
$\tilde A = G_s(\vartheta, \varphi)A$, has the action
\begin{equation}\label{row transform}
\tilde A_{i, \cdot} = \begin{cases} e^{\mathfrak i \varphi}\cos\vartheta A_{s, \cdot} + e^{-\mathfrak i\varphi}\sin\vartheta  A_{s+1, \cdot}, & \mbox{if} \;\; i = s, \\
-e^{\mathfrak i \varphi}\sin\vartheta  A_{s, \cdot} + e^{-\mathfrak i\varphi}\cos\vartheta  A_{s+1, \cdot}, & \mbox{if} \;\; i = s+1,\\
A_{i, \cdot}, & \mbox{otherwise},
\end{cases}
\end{equation}
where $A_{i, \cdot}$ denotes the $i$th row of $A$.

Let us write the matrix elements $A_{s, j}$ and $A_{s+1, j}$, with
$j$ arbitrary but fixed, in polar form, i.e., $A_{s, j} = x
e^{\mathfrak i\phi_x}$ and $A_{s+1, j} = y e^{\mathfrak i\phi_y}$,
with $x,y \geq 0$. We stick to the convention that the phases
$\phi_x$ and $\phi_y$ be in the interval $]{-\pi}, \pi]$ in order to
make this representation unique. It is now easy to see that by
choosing
\begin{gather}
\varphi = (\phi_y - \phi_x)/2, \label{angles1} \\
\vartheta =\arctan\frac{y}{x}, \label{angles2}
\end{gather}
we obtain
\begin{equation}
(G_s(\vartheta, \varphi)A)_{s+1, j} = 0,
\end{equation}
while all the other entries in the $s$th and $(s+1)$th row have
changed according to Eqs. \eqref{row transform}. In the case $x =
0$, we set $\vartheta = \pi/2$ and $\varphi = 0$. In the case $y =
0$, we have $\vartheta = 0$, and we choose to set $\varphi = 0$ as
well. The angles $\vartheta$ and $\varphi$ are thus restricted to
the intervals $\vartheta\in [0, \frac{\pi}{2}]$ and $\varphi \in
]{-\pi}, \pi[$.

By successively applying Givens rotations with appropriately chosen
angles according to Eqs. \eqref{angles1} and \eqref{angles2}, we may
now generate zeroes in $A$ column by column, from left to right,
bottom to top. In greater detail, we first erase the whole first
column, except for the top entry which will generally remain
non-zero. Continuing at the bottom of the second column, we may
generate zeros up to (and including) the third entry from the top of
the column. If we tried to make the second entry zero, we would in
general generate a non-zero entry in the second row of the first
column according to the transformation Eq. \eqref{row transform}. It
is convenient to label the angles calculated during this process by
two indices, and to use the abbreviation $G_s(i, j) =
G_s(\vartheta_{ij}, \varphi_{ij})$. Eventually, we obtain a matrix
$\tilde R$ given by
\begin{equation}\label{givens decompo}
\tilde R = \tilde Q^{-1}A = \prod_{i = 0}^{r - 1}\left[ \prod_{j =
r- i}^{k - 1}G_j(r-i, k - j)\right]A.
\end{equation}
The inner of the two products generates zeros in column $r-i$ from
the bottom up to (and including) row number $r-i + 1$. The upper
block of $\tilde R$ consisting of the first $r$ rows is of upper
triangular form, while the lower block is zero. As a product of
unitary Givens rotations, $\tilde Q^{-1}$ is itself unitary and in
particular invertible. Hence, $\tilde Q$ always exists and is
unitary. We may therefore write
\begin{equation}
A = \tilde Q \tilde R = QR,
\end{equation}
where $Q \in St(k, r)$ consists of the first $r$ columns of $\tilde
Q$ and $R$ is the upper $r\times r$ block of $\tilde R$. Since we
assumed that $A \in St(k, r)$, we have
\begin{equation}
1_{r\times r} = A^\dag A = (QR)^\dag QR = R^\dag Q^\dag QR = R^\dag
R,
\end{equation}
and hence, $R$ is unitary. It is straightforward to see that a
unitary upper triangular matrix can only be of the form
\begin{equation}
R_{ij} = \delta_{ij}e^{\mathfrak{i} \chi_i},
\end{equation}
i.e., a diagonal matrix with only phases on the diagonal. Again, we
may choose $\chi_i \in ]{-\pi}, \pi]$.

We have thus achieved a \textit{unique} parametrization of an
arbitrary matrix $A\in St(k, r)$ by a tuple of Euler-Hurwitz angles
$(\mathbf{\vartheta}, \mathbf{\varphi}, \mathbf{\chi}) \in
\mathfrak{S}$, where
\begin{equation}
\mathfrak{S} = [0, \frac{\pi}{2}]^{r(k - \frac{r+1}{2})}\, \times\,
]{-\pi}, \pi[^{r(k - \frac{r+1}{2})}\, \times \,]{-\pi}, \pi]^r.
\end{equation}
As required, we find that the number of free parameters in this
representation is equal to the dimension of the Stiefel manifold,
i.e., $\dim St(k,r) = 2kr - r^2$. It is clear that the procedure
described above is fully invertible. Hence, we have obtained a
one-to-one mapping $F: \mathfrak{S} \rightarrow St(k, r)$. In
detail, this mapping, for a vector $(\mathbf{\vartheta},
\mathbf{\varphi}, \mathbf{\chi}) \in \mathfrak{S}$, is carried out
by filling an otherwise empty $k\times r$ matrix $B$ with the
entries $B_{ii} = e^{\mathfrak{i}\chi_i}$, $i = 1, \ldots r$. Then,
we apply inverse Givens rotations (specified by the Euler-Hurwitz
angles $\mathbf{\vartheta}$ and $\mathbf{\varphi}$) from the left to
$B$, in inverse order with respect to Eq. \eqref{givens decompo}.

In conclusion, we have transformed the optimization problem Eq.
\eqref{general problem} into the new problem
\begin{equation}\label{parameterized problem}
M(\rho) = \min_{k \geq r}\inf_{s\in \mathfrak{S}} h(F(s)).
\end{equation}
Due to the periodic dependence of $F(s)$ on the angles $s$, it is
practical to expand the search space from $\mathfrak{S}$ to the
whole Euclidean space, making Eq. \eqref{parameterized problem} a
completely unconstrained optimization problem (at the cost of
over-parameterizing the search space \cite{Footnote1}). This problem
can then be solved using standard numerical techniques. In all our
calculations, we have used a quasi-Newton algorithm
\cite{Nocedal1999} together with the line search \texttt{linmin}
mentioned earlier. This method requires first derivatives of the
target function with respect to the angles. The derivatives with
respect to $F$ have already been stated in Eqs. (\ref{partial h 1},
\ref{partial h 2}), and the derivatives of $F$ with respect to the
angles are obtained straightforwardly since each angle appears only
once in the product representation presented above. In order to find
a good approximation to the global minimum, one should restart with
random initial conditions several times and take the over-all
minimum.

\subsection{Test Cases}\label{sec:test cases}

Here, we briefly present some performance results of the two
algorithms presented above. We have applied them to the evaluation
of two different convex-roof entanglement measures for which the
numerical data can be verified by analytically known results.
Although our algorithms show comparatively good performance in these
cases, we would like to stress that the efficiency of a certain
method depends strongly on the type of problem present, and may even
be related to the particular instance of the problem (see the GHZ/W
example below). We have for instance also studied certain matrix
approximation problems, in some of which the parameterized
quasi-Newton method converged very poorly, whereas the modified
steepest descent and the generalized conjugate gradient method were
equally strong and very efficient. One thus cannot generically claim
one algorithm to be better than the other. It is just beneficial to
have several different techniques at hand, out of which one can
choose the best-performing one when applied to a particular given
problem.

\subsubsection{Entanglement of formation of random $2\times 2$
states}

\begin{figure}[t] \includegraphics[width=8cm]{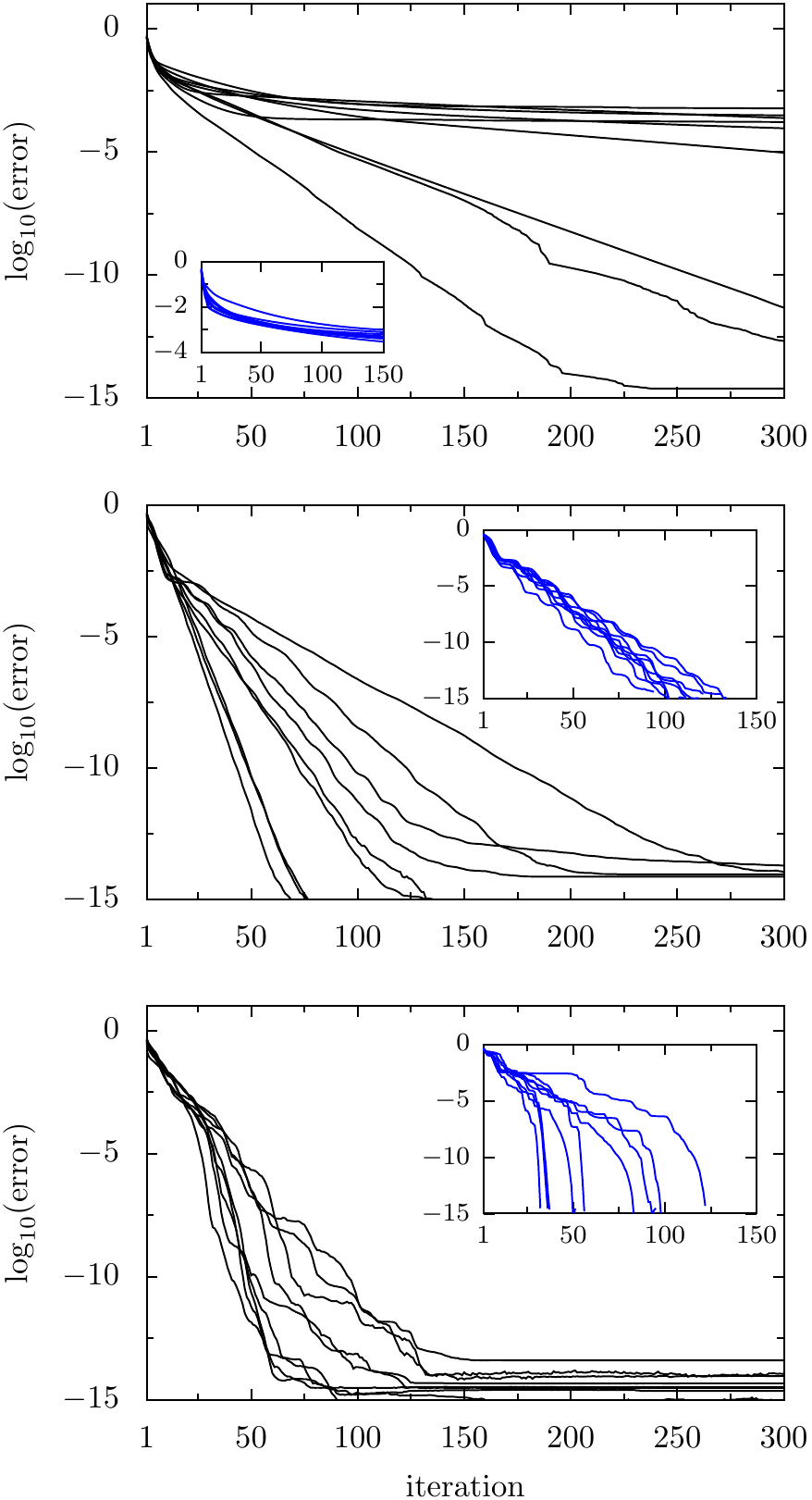}\\
\caption{(color online). Convergence plots of the algorithms used to
evaluate the entanglement of formation on ten random full-rank
two-qubit states (each plot was done using the same ten states)
showing the difference between the numerical data and the analytical
result as a function of the iteration number. Top: The modified
steepest descent algorithm from Ref.~\cite{Manton2002}; Middle: The
generalized conjugate gradient method from Sec.~\ref{sec:cg method};
Bottom: Quasi-Newton on the parameterized search space,
Sec.~\ref{sec:parameterization method}. Each curve in the main plot
is averaged over ten randomly chosen initial points. The typical
behavior of the algorithms for a single fixed density matrix, but
with varying initial points of the iteration, is displayed in the
insets. }\label{fig:eof}
\end{figure}

The entanglement of formation \cite{Bennett1996} is a popular
entanglement measure for bipartite mixed states. It is defined as
the convex roof of the entropy of entanglement \cite{Bennett1996a},
which is, for a state $|\psi\rangle$, the von-Neumann entropy
$S(\rho) = -\trace \rho \log_2 \rho$ of the reduced density matrix
$\rho = \trace_B|\psi\rangle\langle\psi|$, $\trace_B$ denoting the
partial trace over the second subsystem.

Figure \ref{fig:eof} shows the convergence behavior of the
algorithms applied to ten random full-rank two-qubit density
matrices. Displayed is the error at each step of the iteration
between the respective iteration value and the true result. The
latter is known analytically from Ref.~\cite{Wootters1998}.

Compared to the algorithms described here, the modified steepest
descent algorithm due to Ref.~\cite{Manton2002} (top panel) performs
rather poorly. We are aware of the fact that we are comparing here a
steepest descent algorithm with two superlinear algorithms. However,
apart from presenting convergence properties, we would like to point
out that the modified steepest descent algorithm often converges to
imprecise solutions, i.e., it gets stuck in undesirable local
minima. Rather than on the starting point, this phenomenon seems to
depend more on the actual density matrix itself.

The conjugate gradient algorithm due to Ref.~\cite{Audenaert2001}
(middle panel) also shows some dependence on the form of the density
matrix, but always reaches satisfactory accuracy. The results for
the parameterized quasi-Newton method (bottom panel) do not, at
first glance, show the typical fast drop to the solution when close
to a good local minimum. This is due to the effect that changing the
starting point seems to have more influence on the number of
required iterations in the case of the quasi-Newton method (see
insets in Fig.~\ref{fig:eof}). When considering single
(non-averaged) runs of the algorithm, the fast convergence to the
minimum becomes visible. In conclusion, the conjugate gradient and
the parameterized quasi-Newton methods perform best in this case,
the latter even slightly better than the former.

\subsubsection{Tangle of GHZ/W mixtures}

The second test case we present here is concerned with the
evaluation of the tangle of the rank-2 mixed states
\begin{equation}\label{ghz w mixed state}
\rho(\eta) = \eta |\mathrm{GHZ^+_3}\rangle\langle\mathrm{GHZ^+_3}| +
(1 - \eta)|\mathrm{W}\rangle\langle\mathrm{W}|,
\end{equation}
where $|\mathrm{GHZ}^+_3\rangle$ has been defined in Eq. \eqref{GHZ
state generalized},
\begin{equation}\label{W state}
|\mathrm{W}\rangle =
\frac{1}{\sqrt{3}}\big(|{\uparrow\downarrow\downarrow}\rangle +
|{\downarrow\uparrow\downarrow}\rangle +
|{\downarrow\downarrow\uparrow}\rangle\big)
\end{equation}
is the three-qubit W state \cite{Dur2000}, and $0 \leq \eta \leq 1$.
The tangle $\tau_p$ \cite{Coffman2000} is an entanglement measure
for pure states of three qubits and is known to be an entanglement
monotone \cite{Dur2000}. It can hence be generalized to mixed states
by the convex roof construction~\eqref{convex-roof em}. We will
denote the mixed-state tangle by $\tau$, in contrast to the
pure-state version $\tau_p$. The definition of $\tau_p$ reads
\begin{equation}\label{tangle pure}
\tau_p(|\psi\rangle) = 4\ |d_1 - 2d_2 + 4d_3|,
\end{equation}
where
\begin{eqnarray}
  d_1&=& \psi^2_{1}\psi^2_{8} + \psi^2_{2}\psi^2_{7} + \psi^2_{3}\psi^2_{6}+ \psi^2_{5}\psi^2_{4},\\
\nonumber  d_2&=& \psi_{1}\psi_{8}\psi_{4}\psi_{5} +
    \psi_{1}\psi_{8}\psi_{6}\psi_{3}
    + \psi_{1}\psi_{8}\psi_{7}\psi_{2} + \psi_{4}\psi_{5}\psi_{6}\psi_{3}\\
    &&\;\;+ \psi_{4}\psi_{5}\psi_{7}\psi_{2} + \psi_{6}\psi_{3}\psi_{7}\psi_{2},\\
  d_3&=& \psi_{1}\psi_{7}\psi_{6}\psi_{4} + \psi_{8}\psi_{2}\psi_{3}\psi_{5},
\end{eqnarray}
and $\psi_1, \psi_2, \ldots, \psi_8$ denote the components of the
state $|\psi\rangle$ represented in an arbitrary product basis. In
this form, the derivatives of $\tau_p$ with respect to the real and
imaginary parts of the components of $|\psi\rangle$, as required by
the gradient Eqs. (\ref{partial h 1}, \ref{partial h 2}), can be
read off most easily. The tangle takes values between $0$ and $1$
and is maximal for GHZ states. The tangle of the states $\rho(\eta)$
has been studied in Ref.~\cite{Lohmayer2006}, where analytical
expressions as a function of $\eta$ where presented. Particularly,
it was found that the tangle vanishes for all $0 \leq \eta \leq
\eta_0$, where $\eta_0 = \frac{4\sqrt[3]{2}}{3 + 4\sqrt[3]{2}}
\approx 0.6269$, and then continuously increases to unity at $\eta =
1$.

\begin{figure}[t!]
  \includegraphics[width=0.9\columnwidth]{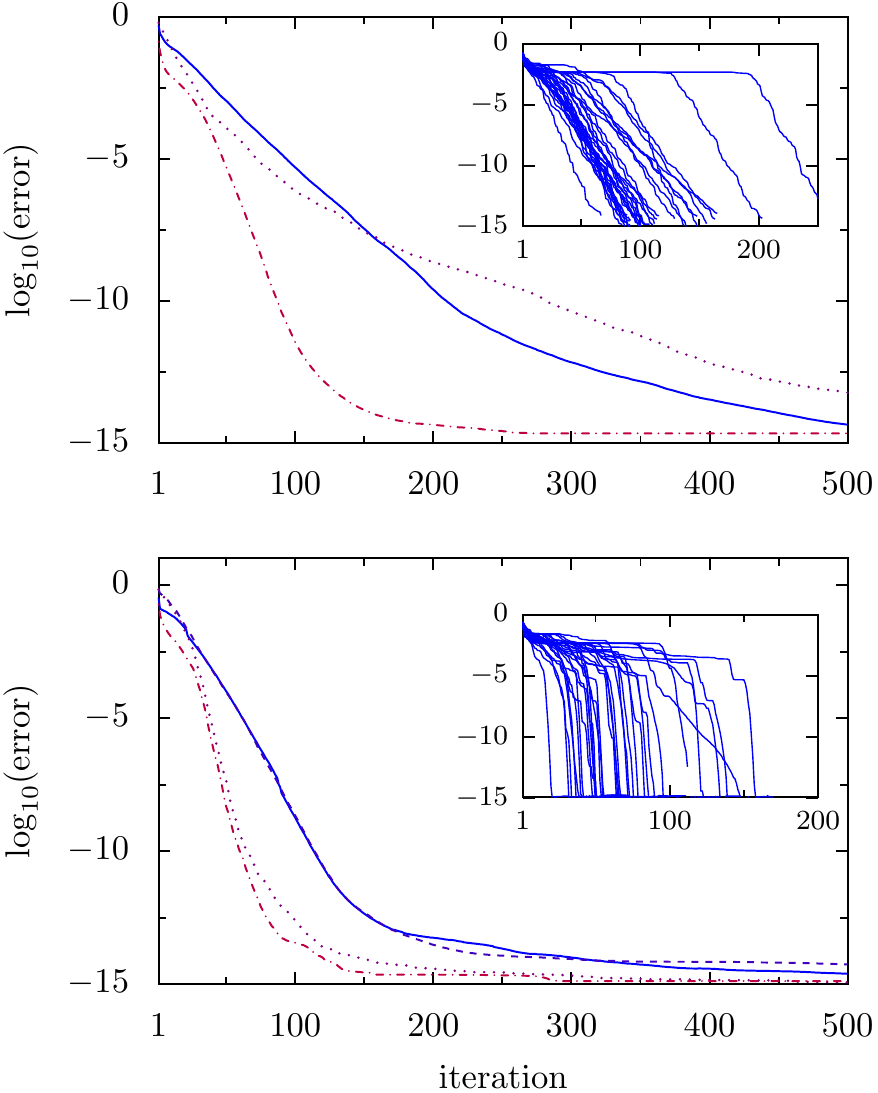}\\
  \caption{(color online). Convergence of the generalized conjugate gradient (top)
and the parameterized quasi-Newton (bottom) algorithms for the
tangle of GHZ/W states Eq. \eqref{ghz w mixed state}. The curves for
the values $\eta = \frac{1}{5}\eta_0$ (solid line), $\eta = (1 -
10^{-4})\eta_0$ (dashed line), $\eta = (1 + 10^{-4})\eta_0$ (dotted
line), and $\eta = \frac{7}{5}\eta_0$ (dashed-dotted line) have each
been obtained by averaging 100 successful runs starting with random
initial points. The typical behavior of single runs is shown in the
insets, where the 100 successful tries that yield one curve in the
main plots are displayed.}\label{fig_ghz-w}
\end{figure}

In Figure \ref{fig_ghz-w} we plot the error between the numerically
obtained and analytically calculated values of $\tau(\rho(\eta))$ as
a function of the iteration number for four particular values of
$\eta$ (see caption of the figure). Only the results of the
generalized conjugate gradient (top panel) and the parameterized
quasi-Newton (bottom panel) method are shown. The modified steepest
descent algorithm from Ref.~\cite{Manton2002} did not succeed to
converge to a reasonable local minimum for the lowest three values
of $\eta$ considered. In these cases, we empirically find the
success rate, which we define as the relative number of final errors
smaller than $10^{-6}$, to be $\lesssim 0.1\%$. For the largest
value of $\eta$ examined, the algorithm showed typical linear
convergence behavior and arrived at a precision around $10^{-12} -
10^{-6}$ after $1000$ iterations with a rather high success rate of
about $60\%$. Similarly, the generalized conjugate gradient
algorithm failed to obtain reasonable results for the value of
$\eta$ slightly below the threshold value $p_0$ in most attempts,
and we find a success rate of $\lesssim 0.2\%$. The success
probability for the other three values of $\eta$ are between $12\%$
and $95\%$, whereas they are between $25\%$ and $80\%$ for the
parameterized quasi-Newton algorithm. One can see, with the help of
looking more detailed into the behavior of single runs (see insets),
that the averaged convergence plots are slightly flattened out due
to some rather rare occurrences of slow convergence. Still, one can
observe that the parameterized quasi-Newton method converges faster
to good local minima.

\subsection{Local unitary equivalence}
We would like to remark here that the parameterized quasi-Newton
method is also capable of determining whether two arbitrary mixed
states are equivalent up to local unitary transformations. While
this problem has an operational solution in some special cases (see,
e.g., Ref. \cite{Fei2005} and references therein), there is no
generally applicable operational criterion known capable of making
this decision. Using the parametrization developed in
Sec.~\ref{sec:parameterization method}, one can express each local
unitary transformation $U_i$ in the matrix $U = U_1\otimes U_2
\otimes \ldots \otimes U_n$ by its Euler-Hurwitz angles and optimize
over the whole set of all angles simultaneously. Furthermore, on can
study in this way how `close' two mixed state are with respect to
local unitary equivalence. Note that such kind of analyses are not
possible with the modified steepest descent or the generalized
conjugate gradient methods, since, as there is no parametrization,
one can optimize over only one unitary matrix at a time.

\section{Physical Application}\label{sec:spin system}

In this section, we use the algorithms developed and described above
to evaluate a multipartite mixed-state entanglement measure of a
concrete physical system.

\subsection{Exchange-coupled spin rings with inhomogeneous magnetic field
geometry}\label{sec:spin rings}

In the following, we consider the Hamiltonian
\begin{equation}\label{Hamiltonian}
H = -J\sum_{i = 1}^N \bsy S_i\cdot \bsy S_{i+1} + b\sum_{i =
1}^N(S_i^x\cos\alpha_i  + S_i^y\sin\alpha_i ),
\end{equation}
where $\bsy S_i = (S_i^x, S_i^y, S_i^z)$, $S_i^k = \sigma^k/2$ with
$\sigma^k$ being the standard Pauli matrices acting on the $i$th
spin, $\bsy S_{N + 1} \equiv \bsy S_1$ and the angles $\alpha_k =
2\pi (k-1)/N$, $k = 1,\ldots, N$. Equation \eqref{Hamiltonian}
describes a closed ring of $N\geq 2$ equidistant exchange-coupled
spin qubits with local in-plane magnetic fields $\bsy b_i \equiv
(b\cos\alpha_i, b\sin\alpha_i, 0)^T$ which are chosen such that the
system is invariant under rotations by multiples of $2\pi/N$ about
the center of the ring. The exchange coupling $J$ is throughout
assumed to be ferromagnetic (i.e., $J > 0$). The fields in Eq.
\eqref{Hamiltonian} are chosen to point radially outwards, but the
following discussion and results also hold for any other local
in-plane field configuration possessing the same rotational
symmetry, since all these systems are local unitary equivalents. The
system is depicted schematically in Fig.~\ref{fig:radial system
3}~{(a)} for three spins.

In fact, we are considering here a generalization of one of the
$N=3$ cases studied in Ref.~\cite{Rothlisberger2008}. There, the
particular field configuration resulted from semiclassical
considerations with the goal of obtaining a state which is close to
a GHZ state [see Eq. \eqref{GHZ state generalized}] as the ground
state of the system. In that case, entanglement can be created by
merely cooling the system to low enough temperatures. In principle,
the argumentation for the occurrence of a GHZ ground state presented
in Ref.~\cite{Rothlisberger2008} can be extended to a number of
qubits $N > 3$. However, it can be expected that for
$N\rightarrow\infty$, the lowest-lying multiplet becomes a
continuous spectrum. Hence, the question arises up to which numbers
of spins $N$ this setup still allows generating GHZ-type
entanglement. Before further investigating this question, we briefly
restate the arguments of Ref.~\cite{Rothlisberger2008} for the
convenience of the reader.

We start from the fact that in the ground state of the classical
analog of the Hamiltonian \eqref{Hamiltonian}, all spins are aligned
for $b = 0$. However, no direction of alignment is favored,
reflecting the full rotational symmetry of the system in spin space.
Small local magnetic fields ($b\ll J$), applied in the way described
above, break this symmetry and one is left with the two degenerate
ground states $\uparrow\uparrow\ldots\uparrow$ and
$\downarrow\downarrow\ldots\downarrow$ where the representation
('quantization') axis is the usual $z$-direction. In fact, each spin
is slightly tilted against its local magnetic field, but there is no
\textit{globally} favored direction of orientation, such as with,
e.g., a global spatially uniform magnetic field. Note that this
effect of tilting vanishes as $b\rightarrow 0$. Due to the Zeeman
term in Eq. \eqref{Hamiltonian} there is an energy barrier between
any path connecting the two degenerate minima. In the quantum case,
tunneling through this barrier lifts the degeneracy between the
ground states and one obtains a tunnel doublet. Thus, in the limit
$b\rightarrow 0^+$, the two lowest lying states are the generalized
GHZ states given in Eq. \eqref{GHZ state generalized}.

\begin{figure}[t]
\includegraphics[width=\columnwidth]{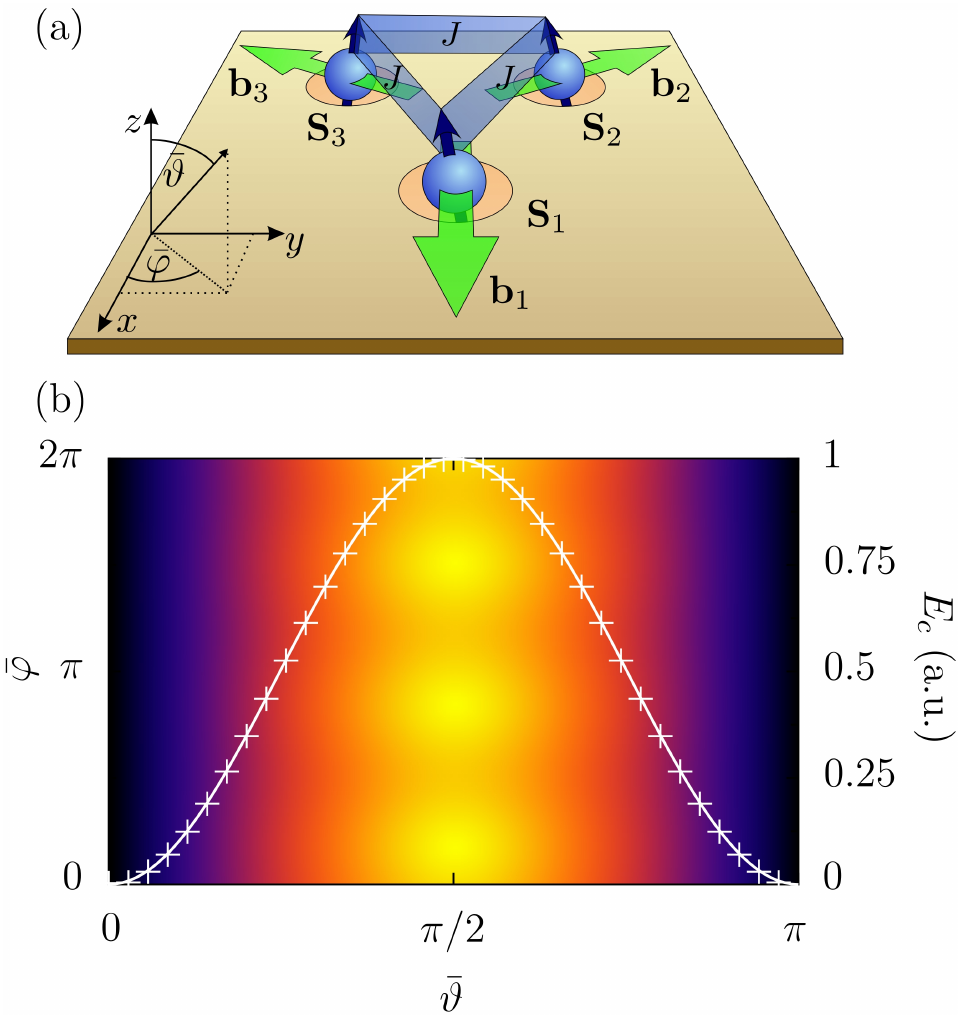}
  \caption{(color online). (a) Schematic depiction of the system described by the Hamiltonian \eqref{Hamiltonian} for $N=3$. Three spins $\bsy S_i$ are situated at the corners of an equilateral triangle
 and are ferromagnetically exchange-coupled with coupling strength $J$. Local radial in-plane magnetic fields $\bsy b_i$ (shown as green arrows in the $xy$-plane) point radially outwards. As discussed in the text,
any other in-plane field geometry obeying the same radial symmetry
(such as, e.g., a `chiral' field looping around the triangle) leads
to equivalent results. (b) Classical energy surface $E_c$ of the
system shown in the top panel. The `mean' angles $\bar\vartheta$ and
$\bar\varphi$ (introduced in the top panel) are well suited to
characterize the state of the system since fluctuations around these
angles are small for $b\ll J$ and sum to zero. The superimposed
white line shows the perturbatively calculated energy barrier at
$\bar\varphi = \pi/2$ [see Eq.~{(2)} in
Ref.~\cite{Rothlisberger2008}], whereas the crosses are due to a
corresponding numerical minimization of the
energy.}\label{fig:radial system 3}
\end{figure}

As an illustration, we plot the energy surface of the classical
three-spin system corresponding to Eq. \eqref{Hamiltonian} in
Fig.~\ref{fig:radial system 3}~{(b)}. We have previously argued [see
Ref.~\cite{Rothlisberger2008}, especially the discussion leading to
Eq.~{(2)} therein] that this energy can be expressed in terms of two
`mean' spherical angles $\bar\varphi$ and $\bar\vartheta$ [cf.
Fig.~\ref{fig:radial system 3}~{(a)}], since all spins will
basically align in the present limit $b\ll J$, up to small
fluctuations which sum to zero and are chosen to minimize the total
energy. One can nicely see how the out-of-plane configurations at
$\bar\vartheta = 0$ and $\bar\vartheta = \pi$ are energetically
favored. For any value of $\bar\varphi$, a path connecting the two
minima has to overcome an energy barrier which scales as $O(b^2)$.
In the figure, this barrier is displayed by the superimposed white
line for the specific value $\bar\varphi = \pi/2$.

Independently of $N$, we are generally confronted with the following
problem if we want to achieve the systems considered here to be in a
highly entangled state at non-zero temperature. On the one hand, the
energy splitting between the ground state and the first excited
state vanishes as $b$ goes to zero. On the other hand, a perfect GHZ
state is obtained exactly in this limit. For increasing magnetic
field, the states continuously deviate from the maximally entangled
GHZ state, as can be imagined with the help of the classical
picture, where the spins start to tilt. One therefore has to choose
the strength of $b$ as a tradeoff between having a highly entangled
ground state and separating this state in energy from the next
higher state.

In order to find this optimal magnetic field strength at a given
temperature $T\neq 0$ we evaluate a suited mixed-state entanglement
measure on the system's canonical density matrix $\rho = \exp(-\beta
H)/\trace\exp(-\beta H)$ where $\beta = 1/\kb T$ and $\kb$ is
Boltzmann's constant. When we studied the case $N = 3$ in
Ref.~\cite{Rothlisberger2008} we used the tangle [see Eq.
\eqref{tangle pure}] as our pure-state measure of choice, since it
is an entanglement measure for three qubits. The generalization to
mixed states was done via the convex-roof construction Eq.
\eqref{convex-roof em}. Here, however, we need a pure-state
entanglement measure which is defined for any $N \geq 2$.

\subsection{Entanglement measure}

In principle, an exponentially increasing number of distinct
entanglement measures is required to capture all possible quantum
correlations in a general pure state of $N$ qudits. This may be
viewed as the reason for the rather large number of proposals for
multipartite entanglement measures that have been put forward over
the last years. Various insights about the structure and
characterization of multipartite entanglement have been gained by
studying such measures. For our purpose, we want to have a measure
that is easy (and fast) to compute (in particular, that is an
analytic function whose complexity grows at most polynomially with
$N$), that captures the type of entanglement present in our system
well, and that possibly has a nice (physical) interpretation. We
found that the Meyer-Wallach measure \cite{Meyer2002}, defined for
an arbitrary number of qubits, fulfills all these criteria.
According to Ref.~\cite{Brennen2003}, it can be written in the
compact form
\begin{equation}\label{Meyer-Wallach}
\gamma(|\psi\rangle) = 2\left[1 - \frac{1}{N}\sum_{k =
1}^N\trace(\rho_k^2)\right],
\end{equation}
where $\rho_k$ is the density matrix obtained by tracing out all but
the $k$th qubit out of $|\psi\rangle\langle\psi|$. This is simply
the subsystem linear entropy averaged over all bipartite partitions
involving one qubit and the rest \cite{Footnote2}. Moreover, it was
shown that this entanglement measure is experimentally observable by
determining a set of parameters that grows linearly with $N$, in
contrast to the exponentially increasing complexity of quantum state
tomography \cite{Brennen2003}. We note at this point that the
Meyer-Wallach entanglement has been generalized to a broader family
of entanglement measures \cite{Scott2004} that might give deeper
insight into the structure of multipartite entanglement. However, we
stick to the simple form \eqref{Meyer-Wallach} for our numerical
calculations, as this measure turns out to describe our type of
entanglement well.

\begin{figure*}[t]
\includegraphics{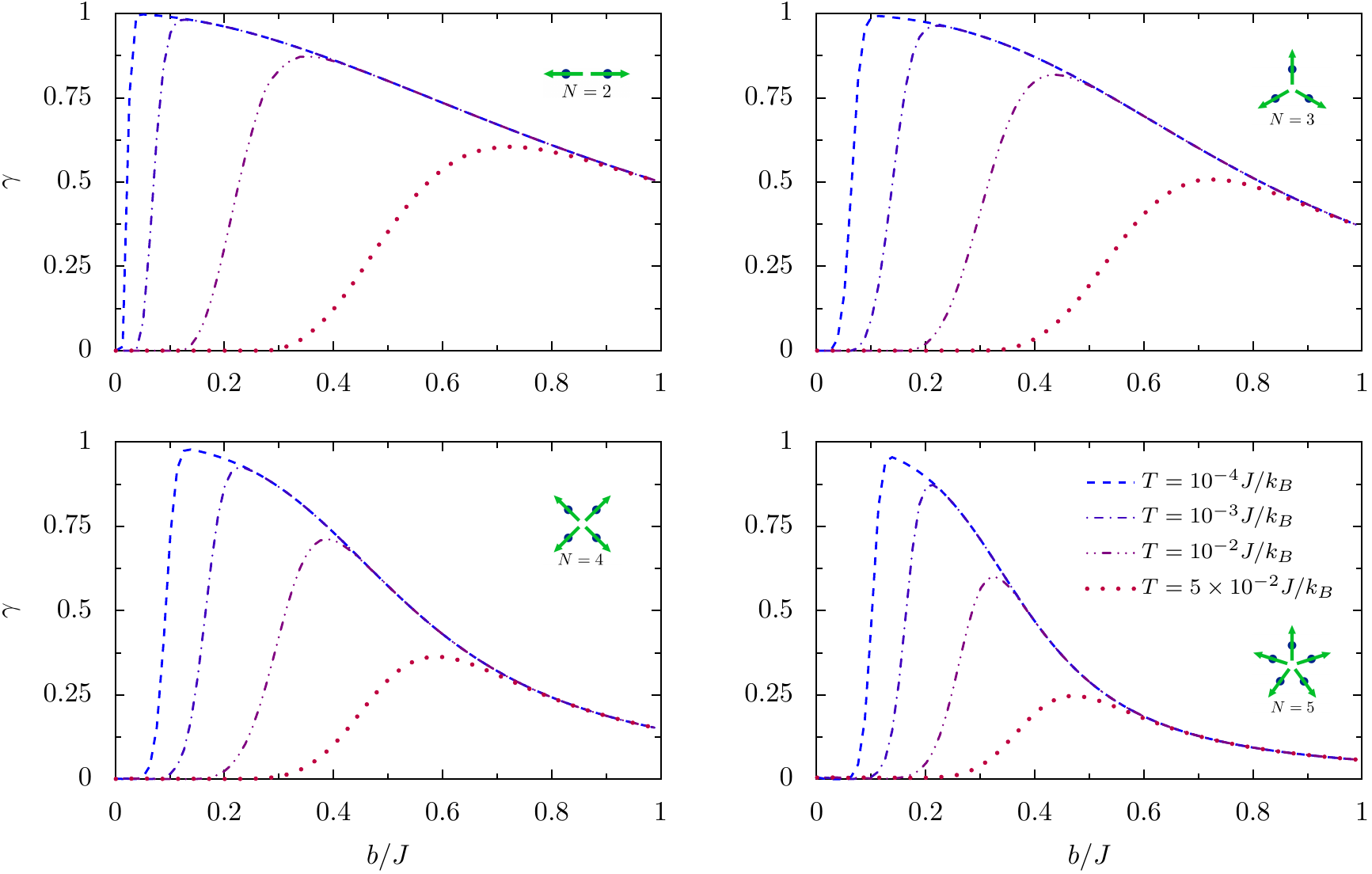}
  \caption{(color online). Meyer-Wallach entanglement measure for the system described by the Hamiltonian Eq. \eqref{Hamiltonian} at different temperatures for several numbers of particles.
Concretely, the cases $N = 2$ (top left), $N = 3$ (top right), $N =
4$ (bottom left), and $N = 5$ (bottom right) are studied at
temperatures $T = 10^{-4} J/\kb$ (dashed line), $T = 10^{-3} J/\kb$
(dashed-dotted line), $T = 10^{-2} J/\kb$ (dashed-dot-dotted line),
and $T = 5\times 10^{-2} J/\kb$ (dotted line).}\label{fig:radial
entanglement}
\end{figure*}

The Meyer-Wallach measure is an entanglement monotone (and can thus
be extended to mixed states via the convex-roof construction), lies
between zero and one, vanishes only for full product states (i.e.,
states of the form $|\psi\rangle = \bigotimes_i|\psi_i\rangle, i =
1,\ldots, N$), and is maximal for generalized GHZ states Eq.
\eqref{GHZ state generalized}. The upper bound is however also
reached by other states, for instance by the so-called cluster
states \cite{Brennen2003, Briegel2001}. A drawback of the
Meyer-Wallach measure is that it can also be maximized by partially
separable states. For example, the state $|\Psi\rangle =
|\Phi\rangle\otimes|\Phi\rangle$, where $|\Phi\rangle =
(|{\uparrow\uparrow}\rangle +
|{\downarrow\downarrow}\rangle)/\sqrt{2}$ is a bipartite Bell state,
gives $\gamma(|\Psi\rangle) = 1$ although it is clearly not globally
entangled \cite{Brennen2003}. This is however not a problem in our
study for two reasons. First of all, we can check by numerical
diagonalization that the ground state of our systems indeed
converges to a multiparite GHZ state (at least for the first few $N
\lesssim 20$). Secondly, comparing the data for $N = 3$ with our
earlier study in Ref.~\cite{Rothlisberger2008} where we had employed
the tangle, we find the same qualitative behavior of both
entanglement measures. Moreover, the optimal values of $b$ for which
the measures reach their maxima at a given temperature coincide
almost perfectly. It is thus reasonable to assume that the
Meyer-Wallach entanglement measure is well suited for quantifying
entanglement in our systems.

The numerical evaluation of the Meyer-Wallach measure extended to
mixed states via the convex-roof construction requires the
derivatives of $\gamma(|\psi\rangle)$ with respect to the real and
imaginary components of $|\psi\rangle$ [see Eqs. (\ref{partial h 1},
\ref{partial h 2})]. Due to the partial traces, these expressions
are a bit cumbersome. However, exploiting the rotational symmetry of
the Hamiltonian studied here, they can be considerably simplified
(see Appendix).

\subsection{Results}\label{sec:results}

Before we present and discuss our numerical results, we would like
to mention that studying the system Eq.~\eqref{Hamiltonian}
analytically for arbitrary $N$ is rather difficult. An exact
diagonalization of the Hamiltonian is not known for arbitrary $N$,
and perturbation theory to constant order in $b$ (independent of
$N$) is not suitable to study the ground-state properties of the
system, since the ground-state splitting is lifted only in $N$-th
order. One can can thus generally expect that the ground-state
splitting scales with the number of spins as $b^N$. Since we must
always have $b \ll 1$, this goes to zero for large $N$, as discussed
in Sec.~\ref{sec:spin rings} above. Obtaining highly entangled
states at finite temperature with this approach will thus be
increasingly difficult for an increasing number of spins $N$.

\begin{figure}[t]
\includegraphics[width = \columnwidth]{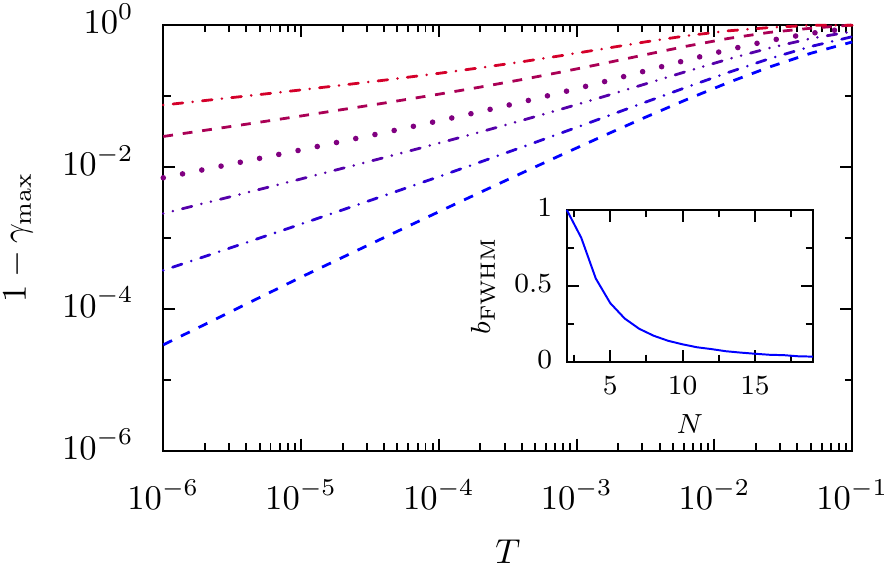}
  \caption{(color online). Difference between zero-temperature and finite-temperature maximally achievable Meyer-Wallach entanglement measure as a function of temperature for
systems with $N = 2,3,4,5,7,10$ spins (from bottom to top). Inset:
Value of the magnetic field strength $b$ as a function of system
size at which the ground state yields the Meyer-Wallach value 0.5
(full width at half maximum).} \label{fig:maximal entanglement}
\end{figure}

Our numerical results are presented in Figs.~\ref{fig:radial
entanglement} and~\ref{fig:maximal entanglement}.
Figure~\ref{fig:radial entanglement} shows the Meyer-Wallach measure
for $N=2,3,4$, and $5$ spins at four different temperatures (see
caption of the figure). Each data point is the result of whichever
of the two algorithms described in Sec.~\ref{sec:algorithms}
performed better in a few trials with random initial conditions.

For a fixed number of spins, the entanglement as a function of the
magnetic field strength $b$ assumes a maximum. This maximal
entanglement $\gamma_\mathrm{max}(T)$ is increased and its position
is shifted to smaller magnetic field values as the temperature is
lowered. This is due to the fact that at low temperatures, only a
small magnetic field is required in order to make the ground-state
splitting sufficiently large compared with temperature. Since these
small field values only slightly disturb the ideal GHZ
configuration, almost maximal values of the entanglement measure
(corresponding to almost perfect GHZ-states) are observed. With
higher temperature, larger field values are required to protect the
ground state. Consistent with the semiclassical picture, this
perturbs the desired spin configuration and leads to a lower amount
of entanglement. For large magnetic fields, all curves coincide
eventually, as the system is always found in the ground state in
that case.

Figure~\ref{fig:maximal entanglement} gives more insight into the
dependence of maximal entanglement $\gamma_\mathrm{max}(T)$ on
temperature and the number of particles. The plot was obtained by
maximizing the Meyer-Wallach measure over the magnetic field
strength $b$ while holding the temperature fixed. Displayed is the
difference between the resulting data to the zero-temperature
maximum (being equal to $1$) as a function of temperature for
different numbers of particles (see caption of the figure). Clearly,
the maximally achievable entanglement $\gamma_\mathrm{max}(T)$
decreases for both increasing temperature and increasing number of
particles. The qualitative dependence on the temperature was
discussed already above. Here we additionally see an almost linear
behavior on a log-log scale at low temperatures, suggesting a
power-law decay of the maximal entanglement of the form $1 -
\gamma_\mathrm{max}(T) \propto T^\alpha$ with an exponent $\alpha$
depending on the number of particles $N$.

The decrease of $\gamma_\mathrm{max}(T)$ with the number of spins
$N$ at fixed temperature is due to the fact that the energy
splitting between the ground and first excited state scales as
$b^N$. With a larger number of particles, a higher magnetic field is
required to achieve a sufficiently large splitting. This in turn
lowers the entanglement in the ground state, due to its
$b$-dependence, resulting in a lowered maximum of the Meyer-Wallach
measure. As an additional obstacle, the ground-state entanglement as
a function of $b$ decays even more rapidly as the number of
particles is increased. This can be seen from the inset of
figure~\ref{fig:maximal entanglement}, where, at $T=0$, the
$b$-values yielding the Meyer-Wallach measure $0.5$ (full width at
half maximum, since the maximum at $T=0$ is always $1$) are shown as
a function of $N$.

\section{conclusions}\label{sec:conclusion}

We have presented two ready-to-use numerical algorithms to evaluate
any generic convex-roof entanglement measure. While one is based on
a conjugate gradient algorithm operating directly on the search
space, the other one is a quasi-Newton procedure performing the
search in the transformed unconstrained Euclidean space. All
required formulas to implement either of the two algorithms have
been stated explicitly, which, in order to calculate different
convex-roof extended pure-state measures, merely leaves the user
with the task of calculating its derivatives with respect to the
real and imaginary components of the pure-state argument. The
relatively different nature of the two procedures increases the
chances that at least one of them performs well in the concrete
application. In a series of numerical tests, we have found that the
algorithms perform well and especially significantly better than
previously presented (non Newton-type) ready-to-use optimization
problems on the Stiefel manifold. However, it is found that the
convergence properties, as is often the case in involved
optimization problems, depend on the cost function. This suggests to
try applying different techniques to a particular optimization
problem and examine which one performs best in that case.

Further, we have applied our algorithms to evaluate a multipartite
entanglement measure on density matrices originating from a real
physical system. The latter consists of $N$ ferromagnetically
exchange-coupled spin-$\frac{1}{2}$ particles placed on the edges of
a regular polygon with $N$ edges. We have argued that a particular
local magnetic field geometry, namely radially symmetric in-plane
fields, favor a highly entangled ground state configuration. We have
confirmed this argumentation by evaluating the mixed-state
Meyer-Wallach entanglement measure, defined for an arbitrary number
of qubits, and found indeed high values of entanglement at low
temperatures and specific magnetic field strengths. This not only
quantifies the entanglement properties present in this system, but
also serves more generally as a proof-of-principle for the
usefulness and applicability of our algorithms.

\acknowledgments

We would like to thank Stefano Chesi and Renato Renner for fruitful
discussions. Financial support from the Swiss NF, the NCCR
Nanoscience, and DARPA Quest is gratefully acknowledged.

\appendix*
\section{Derivatives of the Meyer-Wallach entanglement measure}

Within our numerical framework, the evaluation of the Meyer-Wallach
measure $\gamma(|\psi\rangle)$ [see Eq.~\eqref{Meyer-Wallach}]
requires its partial derivatives with respect to the real and
imaginary components of $|\psi\rangle$ [see Eqs.~(\ref{partial h 1},
\ref{partial h 2})]. They are given by
\begin{widetext}
\begin{align}\label{meyer-wallach derivatives}
  \left.\frac{\partial \gamma}{\partial\re\psi^{(i)}}\right|_{|\psi\rangle} &= -\frac{8}{N}\sum_{k = 1}^N\sum_{\mu = 0}^{1}\sum_{\nu_1 = 0}^{1}\sum_{\nu_2 = 0}^{1}\cdots\sum_{\nu_N = 0}^{1}\re\left(\psi^{(\nu_1,\nu_2,\ldots,\nu_{k-1},\mu,\nu_{k+1},\ldots, \nu_N)}\rho_k^{\nu_k, \mu}\right), \\
  \left.\frac{\partial \gamma}{\partial\im\psi^{(i)}}\right|_{|\psi\rangle} &= -\frac{8}{N}\sum_{k = 1}^N\sum_{\mu = 0}^{1}\sum_{\nu_1 = 0}^{1}\sum_{\nu_2 = 0}^{1}\cdots\sum_{\nu_N = 0}^{1}\im\left(\psi^{(\nu_1,\nu_2,\ldots,\nu_{k-1},\mu,\nu_{k+1},\ldots, \nu_N)}\rho_k^{\nu_k,
\mu}\right).
\end{align}
\end{widetext}
Here, we represented the $i$th component of $|\psi\rangle$ by the
tuple $i = (\nu_1, \ldots, \nu_N)$, with the indices $\nu_j\in\{0,
1\}$ corresponding to some arbitrary product basis
$\{|\nu_1\rangle|\nu_2\rangle\cdots|\nu_N\rangle\}_{\nu_j = 0}^1$ of
the spin system. Furthermore, $\rho_{k}^{\nu, \mu}$ denotes the
matrix element with indices $(\nu, \mu)$ of the reduced density
matrix $\rho_k = \trace_{\nu_1, \nu_2, \ldots, \nu_{k-1}, \nu_{k+1},
\ldots, \nu_N}|\psi\rangle\langle\psi| \in \mathds{C}^{2\times2}$.

In case of the systems studied in Sec.~\ref{sec:spin rings}, the
computation of the Meyer-Wallach measure \eqref{Meyer-Wallach} and
its derivatives can be greatly simplified by exploiting the
rotational symmetry of the Hamiltonian $H$. Since we have $[H,
\mathcal{R}] = 0$, where $\mathcal{R}$ is the symmetry operator for
the rotation by an angle of $2\pi/N$ about the central axis
perpendicular to the plane of the spin ring, all $\rho_k$ in
Eq.~\eqref{Meyer-Wallach} are unitary equivalents for a simultaneous
eigenstate $|\psi\rangle$ of $H$ and $\mathcal{R}$. This reduces Eq.
\eqref{Meyer-Wallach} to the simple form
\begin{equation}
\gamma(|\psi\rangle) = 2[1 - \trace(\rho_1^2)].
\end{equation}
The corresponding derivatives read
\begin{align}
  \left.\frac{\partial \gamma}{\partial\re\psi^{(i)}}\right|_{|\psi\rangle} &= -8[\rho_1^{0,0}\re\psi^{(i)} + \re(\rho_1^{0,1}\psi^{(2^{N-1}+i)})],\\
  \left.\frac{\partial \gamma}{\partial\im\psi^{(i)}}\right|_{|\psi\rangle} &= -8[\rho_1^{0,0}\im\psi^{(i)} +
\im(\rho_1^{0,1}\psi^{(2^{N-1}+i)})],
\end{align}
for $0 \leq i \leq 2^{N-1} - 1$, and
\begin{align}
  \left.\frac{\partial \gamma}{\partial\re\psi^{(i)}}\right|_{|\psi\rangle} &= -8[\rho_1^{1,1}\re\psi^{(i)} + \re(\rho_1^{0,1}\psi^{\ast(i - 2^{N-1})})],\\
  \left.\frac{\partial \gamma}{\partial\im\psi^{(i)}}\right|_{|\psi\rangle} &= -8[\rho_1^{1,1}\im\psi^{(i)} - \im(\rho_1^{0,1}\psi^{\ast(i -
2^{N-1})})],
\end{align}
for $2^{N-1} \leq i \leq 2^N - 1$. In practice, we first diagonalize
$H$ numerically \cite{Footnote3}, subsequently diagonalize further
any degenerate spaces with respect to $\mathcal{R}$, and then apply
the simplified formulas above.


\begin{thebibliography}{43}
\bibitem{Footnote1} Allowing the iteration points to move outside of $\mathfrak{S}$
may result in large absolute values of the angles, possibly leading
to reduced precision in the output. Such effects can be prevented by
projecting the current iteration point back to $\mathfrak{S}$ a few
times in the first couple of iterations.
\bibitem{Footnote3} Since for the number of particles $N$ considered in
Sec.~\ref{sec:results} the splitting between the lowest and the next
higher mutliplet is still always large compared with temperature, we
diagonalize only the lowest-lying $(N+1)$-dimensional subspace of
$H$ using a Lanczos algorithm.
\end{thebibliography}
\end{document}